\newif\ifplainstyle
\plainstyletrue
\plainstylefalse

\newif\ifjhepstyle
\jhepstyletrue

\newif\ifprstyle
\prstyletrue
\prstylefalse

\ifprstyle
	\documentclass[twocolumn,amsmath,amssymb,nofootinbib]{revtex4-1}
\else
	\documentclass[11pt,a4paper]{article}
\fi

\ifjhepstyle
	\usepackage{jheppub}
	\usepackage{amsfonts}
	\usepackage{verbatim}
	\usepackage{float}
	\usepackage{color}
	\usepackage{array}
	\newcolumntype{C}[1]{>{\centering\arraybackslash$}p{#1}<{$}}
	\makeatletter
	\def\@fpheader{\phantom{Prepared for submission to JHEP}}
	\makeatother
\else	
 	\ifprstyle
		\usepackage{verbatim}
		\usepackage{amsmath,amsfonts,amssymb}
		\usepackage[colorlinks=true
                	,urlcolor=blue
                	,anchorcolor=blue
                	,citecolor=blue
                	,filecolor=blue
                	,linkcolor=blue
                	,menucolor=blue
                	]{hyperref}
	\else
            	\usepackage{verbatim}
            	\usepackage{cite}
            	\usepackage{setspace}
            	\usepackage[top=2.5cm, bottom=2.75cm, left=2.5cm, right=2.5cm]{geometry}
            	\usepackage{amsmath,amsfonts,amssymb}
            	\usepackage[colorlinks=true
            	,urlcolor=blue
            	,anchorcolor=blue
            	,citecolor=blue
            	,filecolor=blue
            	,linkcolor=blue
            	,menucolor=blue
            	,linktoc=page
            	]{hyperref}
            	\usepackage{float}
            	\restylefloat{table}
            	
            	\numberwithin{equation}{section}
	\fi
\fi

\allowdisplaybreaks

\usepackage{tikz}
\usepackage[font=small]{caption}

\newcommand{\ThisIsTheTitle}{Non-linear partially massless symmetry in an SO(1,5) continuation of conformal gravity} 
\newcommand{\ThisIsAuthorOne}{Luis Apolo}
\newcommand{\ThisIsEmailOne}{luis.apolo@fysik.su.se}
\newcommand{\ThisIsAuthorTwo}{and S.~F. Hassan}
\newcommand{\ThisIsEmailTwo}{fawad@fysik.su.se}
\newcommand{\ThisIsTheAffiliation}{Department of Physics \& The Oskar Klein Centre, \\
Stockholm University, AlbaNova University Centre, SE-106 91 Stockholm, Sweden}
\newcommand{\TheseAreTheKeywords}{}

\newcommand{\ThisIsTheAbstract}{We construct a non-linear theory of interacting spin-2 fields that is invariant under the partially massless (PM) symmetry to all orders. This theory is based on the $SO(1,5)$ group, in analogy with the $SO(2,4)$ formulation of conformal gravity, but has a quadratic spectrum free of ghost instabilities. The action contains a vector field associated to a local $SO(2)$ symmetry which is manifest in the vielbein formulation of the theory.  We show that, in a perturbative expansion, the $SO(2)$ symmetry transmutes into the PM transformations of a massive spin-2 field. In this context, the vector field is crucial to circumvent earlier obstructions to an order-by-order construction of PM symmetry. Although the non-linear theory lacks enough first class constraints to remove all helicity-0 modes from the spectrum, the PM transformations survive to all orders. The absence of ghosts and strong coupling effects at the non-linear level are not addressed here.}

\ifjhepstyle
\title{\ThisIsTheTitle}

\author{\ThisIsAuthorOne}
\author{\ThisIsAuthorTwo}

\affiliation{\ThisIsTheAffiliation}

\emailAdd{\ThisIsEmailOne}
\emailAdd{\ThisIsEmailTwo}

\abstract{\ThisIsTheAbstract} 

\fi

\begin{document}

\ifjhepstyle
\maketitle
\flushbottom
\fi

\long\def\symfootnote[#1]#2{\begingroup%
\def\thefootnote{\fnsymbol{footnote}}\footnote[#1]{#2}\endgroup} 

\def\rednote#1{{\color{red} #1}}
\def\bluenote#1{{\color{blue} #1}}

\def\({\left (}
\def\){\right )}
\def\lb{\left [}
\def\rb{\right ]}
\def\lB{\left \{}
\def\rB{\right \}}

\def\Int#1#2{\int \textrm{d}^{#1} x \sqrt{|#2|}}
\def\Bra#1{\left\langle#1\right|} 
\def\Ket#1{\left|#1\right\rangle}
\def\BraKet#1#2{\left\langle#1|#2\right\rangle} 
\def\Vev#1{\left\langle#1\right\rangle}
\def\Vevm#1{\left\langle \Phi |#1| \Phi \right\rangle}\def\bbox{\bar{\Box}}
\def\til#1{\tilde{#1}}
\def\wtil#1{\widetilde{#1}}
\def\ph#1{\phantom{#1}}

\def\ra{\rightarrow}
\def\la{\leftarrow}
\def\lra{\leftrightarrow}
\def\p{\partial}
\def\diff{\mathrm{d}}

\def\sinh{\mathrm{sinh}}
\def\cosh{\mathrm{cosh}}
\def\tanh{\mathrm{tanh}}
\def\coth{\mathrm{coth}}
\def\sech{\mathrm{sech}}
\def\csch{\mathrm{csch}}

\def\a{\alpha}
\def\b{\beta}
\def\g{\gamma}
\def\d{\delta}
\def\e{\epsilon}
\def\ve{\varepsilon}
\def\k{\kappa}
\def\l{\lambda}
\def\n{\nabla}
\def\om{\omega}
\def\s{\sigma}
\def\t{\theta}
\def\z{\zeta}
\def\vp{\varphi}

\def\ss{\Sigma}
\def\dd{\Delta}
\def\gg{\Gamma}
\def\ll{\Lambda}
\def\tt{\Theta}

\def\A{{\cal A}}
\def\B{{\cal B}}
\def\cE{{\cal E}}
\def\D{{\cal D}}
\def\F{{\cal F}}
\def\H{{\cal H}}
\def\I{{\cal I}}
\def\J{{\cal J}}
\def\K{{\cal K}}
\def\L{{\cal L}}
\def\O{{\cal O}}
\def\P{{\cal P}}
\def\cS{{\cal S}}
\def\W{{\cal W}}
\def\X{{\cal X}}
\def\Z{{\cal Z}}

\def\we{\wedge}
\def\re{\textrm{Re}}

\def\tilw{\tilde{w}}
\def\tile{\tilde{e}}

\def\zz{\bar z}
\def\xx{\bar x}
\def\xp{x^{+}}
\def\xm{x^{-}}

\def\VirU1{\mathrm{Vir}\otimes\hat{\mathrm{U}}(1)}
\def\VirSL2R{\mathrm{Vir}\otimes\widehat{\mathrm{SL}}(2,\mathbb{R})}
\def\U1{\hat{\mathrm{U}}(1)}
\def\SL2R{\widehat{\mathrm{SL}}(2,\mathbb{R})}
\def\sl2r{\mathrm{SL}(2,\mathbb{R})}
\def\by{\mathrm{BY}}

\def\RR{\mathbb{R}}

\def\tr{\mathrm{tr}}
\def\bnabla{\overline{\nabla}}

\def\sint{\int_{\ss}}
\def\dsint{\int_{\p\ss}}
\def\hint{\int_{H}}

\newcommand{\eq}[1]{\begin{align}#1\end{align}}
\newcommand{\eqst}[1]{\begin{align*}#1\end{align*}}
\newcommand{\eqsp}[1]{\begin{equation}\begin{split}#1\end{split}\end{equation}}

\newcommand{\absq}[1]{{\textstyle\sqrt{\left |#1\right |}}}

\ifprstyle
\title{\ThisIsTheTitle}

\author{\ThisIsAuthorOne}
\email{\ThisIsEmailOne}

\author{\ThisIsAuthorTwo}
\email{\ThisIsEmailTwo}

\affiliation{\ThisIsTheAffiliation}


\begin{abstract}
\ThisIsTheAbstract
\end{abstract}


\maketitle

\fi

\ifplainstyle
\begin{titlepage}
\begin{center}

\ph{.}

\vskip 4 cm

{\Large \bf \ThisIsTheTitle}

\vskip 1 cm

\renewcommand*{\thefootnote}{\fnsymbol{footnote}}

{{\ThisIsAuthorOne}\footnote{\ThisIsEmailOne} and {\ThisIsAuthorTwo}\footnote{\ThisIsEmailTwo}}

\renewcommand*{\thefootnote}{\arabic{footnote}}

\setcounter{footnote}{0}

\vskip .75 cm

{\em \ThisIsTheAffiliation}

\end{center}

\vskip 1.25 cm

\begin{abstract}
\noindent \ThisIsTheAbstract
\end{abstract}

\end{titlepage}

\newpage

\fi

\ifplainstyle
\tableofcontents
\noindent\hrulefill
\bigskip
\fi

\section{Introduction}
\label{se:intro}

It is well known that local gauge symmetries are crucial for the
consistency of the Standard Model of particle physics by ensuring its
unitarity and renormalizability. On the contrary, the Einstein-Hilbert
action does not admit additional gauge symmetries besides the usual
diffeomorphisms. Hence it is natural to search for extensions of
General Relativity with extra gauge symmetries in the hope of
improving the quantum behaviour of the theory. Two such theories that 
have attracted considerable attention are conformal gravity, which is
invariant under local scale transformations, and partially massless
gravity (see below for details). However, while a nonlinear action for
conformal gravity is known, it contains higher-order time derivatives and therefore an Ostrogradsky ghost instability. On the
other hand, a theory of spin-2 fields with a ``partially massless''
gauge symmetry is not known to exist beyond the linear level. The goal
of this paper is to construct a nonlinear theory with partially
massless symmetry that is also related to conformal gravity but
avoids its ghost instability, at least to linear order.

In more detail, let us recall that spin-2 fields in de Sitter space
fall into one of three representations of its isometry group
corresponding to massless, massive, or partially massless (PM)
fields~\cite{Deser:1983mm,
Deser:2001pe,Deser:2001us,Deser:2001wx,Deser:2001xr,Zinoviev:2001dt,Deser:2004ji}.
The latter are described by the linear Fierz-Pauli
theory~\cite{Fierz:1939ix} where the mass of the spin-2 field
saturates the Higuchi bound~\cite{Higuchi:1986py},
  \eq{
  m^2 = \frac{2}{d-1} \ll. \label{higuchi}
  }
Here $\ll$ is the cosmological constant and $d$ is the dimension of
spacetime. At this point in parameter space a new gauge symmetry
emerges where the spin-2 field $\vp_{\mu\nu}$ transforms as 
  \eq{
  \d \vp_{\mu\nu} = \( \overline{\nabla}_{\mu} \overline{\nabla}_{\nu}
  + \frac{2}{(d-2)(d-1)} \ll\, \bar{g}_{\mu\nu} \) \xi(x), \label{pmsymmetry}  
  }
and $\bar{g}_{\mu\nu}$ is the de Sitter metric. This local
symmetry is responsible for removing the helicity-0 component of
$\vp_{\mu\nu}$.  Thus, in four dimensions partially massless fields propagate 4 instead of the 5 degrees of freedom that characterize a massive spin-2 field. Going beyond the linear equations, one may ask if an interacting theory exists that is invariant under a generalization of the above transformation.

In general, non-linear ghost-free theories exist for interacting
massless and massive spin-2 fields -- namely General Relativity,
massive~\cite{deRham:2010kj,Hassan:2011vm,Hassan:2011hr,Hassan:2011tf},
and multi-metric
gravity~\cite{Hassan:2011zd,Hassan:2011ea,Hinterbichler:2012cn} (for
reviews see~\cite{Hinterbichler:2011tt, deRham:2014zqa,
Schmidt-May:2015vnx}). However, these theories do not admit extra
gauge symmetries and no unitary, non-linear theory of partially massless fields
is known.  Although specific subclasses of massive~\cite{deRham:2012kf},
bimetric~\cite{Hassan:2012gz,Hassan:2012rq,Hassan:2013pca,Hassan:2015tba}
and multimetric gravity~\cite{Baldacchino:2016jsz} exhibit
interesting PM features, several no-go results suggest that
non-linear theories of PM fields
containing at most two derivatives do not
exist~\cite{Zinoviev:2006im,Deser:2013uy,deRham:2013wv,
Garcia-Saenz:2014cwa,Garcia-Saenz:2015mqi,Apolo:2016vkn}. In
particular, in the massive gravity approach -- with a single PM field
in a fixed background -- an order by order construction in powers of
$\vp_{\mu\nu}$ encounters obstructions that prevent extension of the
PM symmetry~\eqref{pmsymmetry} beyond cubic
order~\cite{Zinoviev:2006im,deRham:2013wv}. A possible way to
circumvent this obstruction is to enlarge the spectrum of the theory
with an additional massless spin-2 field that transforms non-trivially
under the PM symmetry, as in the bimetric approach. However, as shown in~\cite{Joung:2014aba,Apolo:2016vkn}, this additional spin-2 field is not sufficient to circumvent the aforementioned obstruction and the PM
symmetry cannot be extended beyond terms that are
cubic in $\vp_{\mu\nu}$. This result suggests that additional fields of lower or higher spin are necessary. 

In this paper we show that the obstruction to non-linear PM symmetry can be circumvented in the presence of an additional vector field. Indeed, starting from a non-linear action and expanding it in powers of $\vp_{\mu\nu}$, we show that the PM symmetry can be
extended to all orders in the fields. The construction of this theory is motivated by conformal (Weyl) gravity, a non-linear theory featuring both massless and
partially massless fields~\cite{Maldacena:2011mk}. In conformal gravity the kinetic term of the PM field comes with the wrong sign, which reflects the
higher-derivative, non-unitary nature of the theory. A naive
analytic continuation of the PM fields $\vp_{\mu\nu}\rightarrow
i\vp_{\mu\nu}$ will render the quadratic theory ghost free, but will
introduce imaginary couplings at odd orders. Given the continued
interest in conformal gravity~\cite{Mannheim:2011ds,Hooft:2014daa},
and its close relationship to partial
masslessness~\cite{Maldacena:2011mk,Deser:2012qg,Joung:2014aba,Hassan:2015tba},
it proves useful to make the notion of such an ``analytic continuation'' 
more precise. For this we recall that
in~\cite{Kaku:1977pa}, conformal gravity was constructed as a gauge
theory of the conformal group $SO(2,4)$, mirroring a similar
construction of Einstein gravity based on
$SO(1,4)$~\cite{MacDowell:1977jt}. In this approach the spin-2 ghost
modes are correlated with the $(2,4)$ signature of the group manifold,
suggesting that a similar construction based on the $SO(1,5)$ group
could lead to a better behaved theory. Interestingly, it has been
pointed out in~\cite{Joung:2014aba} that, perturbatively, candidate bimetric PM theories exhibit an $SO(1,5)$ global
symmetry.

Motivated by these considerations, our starting point in this paper is
a gravity action based on the $SO(1,5)$ group.\footnote{Our theory is
the simplest in a family of four-dimensional theories of charged
gravity in de Sitter space. The latter may be realized as gauge
theories of $SO(1,3) \times SO(n) \subset SO(1,3+n)$ mirroring similar
constructions for Einstein gravity based on
$SO(1,4)$~\cite{MacDowell:1977jt}, and conformal gravity based on
$SO(2,4)$~\cite{Kaku:1977pa}. Note that analogous constructions exist
for Einstein and conformal gravity in three
dimensions~\cite{Achucarro:1987vz,Witten:1988hc,Horne:1988jf}. Our
approach is also related to that of
refs.~\cite{Gwak:2015vfb,Gwak:2015jdo} that construct
three-dimensional theories of colored gravity featuring partially
massless fields at linearized order.} Our approach is closely related
to that of conformal gravity~\cite{Kaku:1977pa}, but supplemented by
additional elements required by symmetries.  Although these elements
seem to originate from an extension of the $\mathfrak{so}(1,5)$
algebra, in this work we keep to the minimal setup. The outcome is a
two-derivative theory that avoids the earlier no-go results and
realizes the partially massless symmetry to all orders. While the
absence of ghosts at the non-linear level and strong coupling issues
are not addressed here, we show that the theory is ghost free at the
linear level. Furthermore, we point out that similar kinetic terms
have been shown to feature nonlinear constraints that remove
propagating modes beyond linear order~\cite{Li:2015izu,Li:2015iwc}.
Hence it is possible that the $SO(1,5)$ theory remains ghost free to
all orders.

In more detail, the $SO(1,5)$ theory contains two vielbeins $e^a{}_\mu$
and $t^a{}_\mu$, a Lorentz spin connection $\omega^{ab}_\mu$, and a
gauge field $A_\mu$. The vielbeins lead to two metrics invariant under local
Lorentz transformations whose perturbations correspond to linear combinations of a massless and a massive spin-2 fields -- the minimum field content required for a PM theory coupled to gravity. Generating dynamics without breaking symmetries
of the action requires promoting $\omega^{ab}_\mu$ to the spin
connection of the complexified Lorentz algebra. 
Altogether these fields lead to a bimetric theory with non-standard kinetic
terms and the specific bimetric potential considered
in~\cite{Hassan:2012gz,Hassan:2012rq} in connection with PM symmetry. 

The key feature of the $SO(1,5)$ theory is the additional vector field
and its local $SO(2) = U(1)$ symmetry under which the two vielbeins of
the theory are charged. We show that, when re-expressed in terms of
canonical spin-2 fields,
the gauge transformations of the vector field and the vielbeins are transmuted
into the partially massless symmetry of a massive spin-2 field.  At
linear order the theory describes massless and partially massless spin-2 fields, as well as a massless vector field, along
with their respective gauge symmetries, denoted here by
$\textrm{Diff}\times\textrm{PM}\times U(1)$. In contrast to conformal gravity, the partially massless field is not a ghost. Furthermore,
while the $\textrm{Diff}$ symmetry is present to all
orders, \emph{only the diagonal $U(1)$ part of the $\textrm{PM}\times
U(1)$ gauge symmetry is present non-linearly}, a fact that is manifest in the
vielbein formulation of the theory. This means that the theory loses one of the
first class constraints present at linear order, which suggests that
not all of the helicity-0 modes decouple.

We call this theory ``partially massless'' in the sense that it circumvents earlier no-go results and yields an action that is invariant under PM transformations~\eqref{pmsymmetry} to all orders. Consequently, the relation~\eqref{higuchi}
between the mass of the spin-2 field and the cosmological constant is
preserved. While we succeed in extending the PM
symmetry to all orders, the massive spin-2 field does not necessarily carry 4 polarizations.\footnote{This may be justified on general grounds:
since spin-2 fields with 4 helicities arise only in de Sitter backgrounds, it
is expected that a non-linear background independent theory must
accommodate massive spin-2 fields (represented as rank-2 symmetric
tensors) with 5 polarizations.} This does not mean that the partially
massless symmetry is trivially realized, e.g.~as in the St\"uckelberg trick. To the contrary, this is how the massive field {\it must} transform in order to render the theory
invariant under local $SO(2)$ transformations. It
is natural to conjecture that the partially massless symmetry is a
generic feature of theories with charged metrics/vielbeins. 
Indeed, a similar phenomenon is observed in three dimensions in
refs.~\cite{Gwak:2015vfb,Gwak:2015jdo}. 

The paper is organized as follows. In Section~\ref{se:so15} we
construct a gauge theory for the $SO(1,3)$ subgroup of $SO(1,5)$ that admits an additional local $SO(2)$ symmetry. In particular, we discuss
the constraints that must be obeyed by the vielbeins and the spin
connection of the theory. In Section~\ref{se:pmsymmetry} we consider
the perturbative metric formulation of the theory up to quadratic order in the
massive spin-2 field. Therein we show how the local $SO(2)$ symmetry
of the vielbeins is transmuted into the partially massless symmetry of
the massive graviton. We end with our conclusions and outlook in
Section~\ref{se:conclusions} where we also comment on the non-abelian
generalization of the partially massless symmetry. In
Appendix~\ref{enhancedalgebra} we present a generalization of the
$\mathfrak{so}(1,5)$ algebra that makes our construction more systematic.

\section{A gauge theory based on $SO(1,5)$}
\label{se:so15}

In this section we construct a gauge theory for $SO(1,3) \times
SO(2) \subset SO(1,5)$. We begin by identifying the appropriate gauge
fields, their transformations properties, and their associated field
strengths. We then propose an action and the additional constraints
necessary to recover the partially massless symmetry in the metric
formulation of the theory.


\subsection{Fields and curvatures} \label{fieldsANDcurvatures}

Our starting point is the $SO(1,5)$ group. This group is the global
symmetry group of the linear theory of a massless and a partially
massless spin-2 fields~\cite{Joung:2014aba}. It is closely related to
the conformal group, $SO(2,4)$, which is the global symmetry group of
conformal gravity. Since conformal gravity can be obtained as a gauge theory
based on $SO(2,4)$, it is natural to ask whether a similar
construction leads to a non-linear theory with partially massless
symmetry. While this turns out to be the case, the analogy is not a
perfect one since our theory does not admit a global $SO(1,5)$
symmetry at non-linear order. What singles out $SO(1,5)$ is that it
admits the direct product of the Lorentz and $SO(2)$ groups as a
subgroup. While the former is characteristic of the vielbein
formulation of gravitational theories, the latter will be important in
the realization of the partially massless symmetry to all orders.

The generators $J_{AB}$ of the $\mathfrak{so}(1,5)$ algebra are characterized by the commutation relations,
  \eq{
  [J_{AB},J_{CD}] = \eta_{AD} J_{BC} + \eta_{BC} J_{AD} - \eta_{AC} J_{BD} - \eta_{BD} J_{AC}, \label{so15algebra}
  }
where $A,B \in \{ 0, \dots, 5\}$ and $\eta_{AB}$ is the Minkowski
metric with signature $(-,+,+,+,+,+)$. It will be convenient to work
with a basis where the $\mathfrak{so}(1,3) \oplus \mathfrak{so}(2)$ subalgebra is
manifest. Introducing the following notation,
  \eq{
  \qquad P^{(1)}_{a} = J_{a4}, \qquad P^{(2)}_{a} = J_{a5}, \qquad D = J_{45},
  }
where $a,b \in \{0, \dots, 3\}$, the non-vanishing commutators of the $\mathfrak{so}(1,5)$ algebra read,
  \eq{
  [J_{ab}, J_{cd}] & = \eta_{ad} J_{bc} + \eta_{bc}J_{ad} - \eta_{ac} J_{bd} - \eta_{bd}J_{ac}, \label{so15algebra1} \\
  [J_{ab}, P^{(i)}_c] & = \eta_{bc} P^{(i)}_a - \eta_{ac} P^{(i)}_b, \\
  [P^{(i)}_a, P^{(j)}_b] & = \e^{ij} \eta_{ab} D - \d^{ij} J_{ab}, \\
  [D, P^{(i)}_a] & = \e^{ij} P_{(j)a}, \label{so15algebra4}
  }
where $i,j \in \{1, 2 \}$, $\d_{ij}$ is a Euclidean metric, and
$\e^{12} = -\e^{21} = -1$. Here the $a, b$ indices are naturally
interpreted as tangent space indices in the vielbein formulation of a
gravitational theory while the $i, j$ indices label vectors of
$SO(2)$.

Following Kaku,
Townsend, and Nieuwenhuizen~\cite{Kaku:1977pa} (see
also~\cite{MacDowell:1977jt,Ne'eman:1978nn,Ivanov:1981wn,Trujillo:2013saa}) we parametrize the $SO(1,5)$ gauge field by the following
one-form,
  \eq{
  \mathbb{A} = \frac{1}{2} \om^{ab} J_{ab} + \ell^{-1} e^a P^{(1)}_a + \ell^{-1} t^a P^{(2)}_a + A D. \label{gaugefield}
  }
Here $\om^{ab}$ and $A$ have dimensions of energy while $e^a$ and
$t^a$ are dimensionless, which explains the presence of the length
scale $\ell$. In order to see that $\om^{ab}$ plays the role of the
spin connection, while $e^a$ and $t^a$ play the role of vielbeins, let
us consider their behaviour under infinitesimal $SO(1,5)$
transformations.\footnote{To simplify the notation we express most of
our equations using forms, e.g.~$e^a = e^a{}_{\mu} d x^{\mu}$ and $t^a =
t^a{}_{\mu} dx^{\mu}$ denote vielbein one-forms. However, we will refer
to $e^a$ and $t^a$ simply as vielbeins.} If we parametrize the latter
by the scalar,
  \eq{
  \l = \frac{1}{2} \ll^{ab} J_{ab} + \chi^a P^{(1)}_a + \z^a P^{(2)}_a + \xi D, \label{gaugetransformation}
  }
then, under infinitesimal gauge transformations of the form
$\d_{\l} \mathbb{A} = d \l + [\mathbb{A},\l]$, we find,
  \eq{
  \d_{\l} \om^{ab} &= \D_{\om} \ll^{ab} + 2 \ell^{-1} \chi^{[a} e^{b]} + 2 \ell^{-1} \z^{[a} t^{b]}, \label{deltaomega} \\
  \d_{\l} e^a &= -\ll^a{}_{b} e^b - \xi\, t^a + \ell \D_{\om}  \chi^a + \ell \z^a A , \label{deltae}  \\
  \d_{\l} t^a &= -\ll^a{}_b t^b +  \xi\, e^a + \ell \D_{\om} \z^a - \ell \chi^a A , \label{deltat}  \\
  \d_{\l} A &=  d \xi + \ell^{-1} \chi_a t^a - \ell^{-1} \z_a e^a. \label{deltaa}
  }
In these equations we (anti)symmetrize indices with unit weight,
e.g.~$\z^{[a}e^{b]} = \tfrac{1}{2} ( \z^{a}e^{b}-\z^{b}e^{a})$ and
$\D_{\om}$ denotes the covariant derivative with respect to the spin
connection $\om^{ab}$, e.g.~$\D_{\om} \z^a = d \z^a + \om^a{}_b \z^b$.

From eq.~\eqref{deltaomega} we see that, as expected, $\om^{ab}$
transforms as the spin connection under Lorentz
transformations. Furthermore it is left invariant under the $SO(2)$
transformations generated by $D$. On the the other hand, from
eqs.~\eqref{deltae} and~\eqref{deltat} we see that both $e^a$ and
$t^a$ transform homogeneously under Lorentz transformations which
motivates their identification as vielbeins. We can also see that
these vielbeins form a vector under $SO(2)$ transformations, i.e.,
  \eq{
  \d_{\xi} \(\begin{array}{c} e^a \\  t^a \end{array}\) = \(\begin{array}{cc} 0 & -\xi \\ \xi & 0 \end{array}\)\(\begin{array}{c} e^a \\  t^a \end{array}\). \label{vielbeintransformation}
  }
In particular, from the $e^a$ and $t^a$ vielbeins we can define the
following charged metrics which are invariant under local Lorentz
transformations,\footnote{The presence of two sets of vielbeins and two metrics suggests that the theory describes two interacting spin-2 degrees of freedom, as desired. Whether these fields are propagating, i.e.~weakly coupled on the de Sitter background, depends on the choice of spin connection as discussed in detail in the following section.}
  \eq{
  g_{\mu\nu} = e^a{}_{\mu} e_{a\nu}, \qquad f_{\mu\nu} = t^a{}_{\mu} t_{a\nu }. \label{fgmetrics}
  }
We can also define a metric that is invariant under both local
Lorentz and $SO(2)$ transformations, namely,
  \eq{
  G_{\mu\nu} = e^a{}_{\mu} e_{a\nu} +  t^a{}_{\mu} t_{a\nu} . \label{so2metric}
  }
Note that $A$ transforms as an $SO(2)$ gauge field and as a Lorentz
scalar, cf.~eq.~\eqref{deltaa}. Thus, the $SO(1,3) \times SO(2)$ subgroup of $SO(1,5)$ forms \emph{a maximal set of symmetries
for which the basic ingredients of the theory, namely the spin
connection, the vielbeins, and the vector, transform appropriately,
i.e.~either as tensors or connections.} For this reason, our theory will be manifestly invariant only under this subgroup.

Having identified the basic fields, let us now consider the
curvatures, or field strengths, from which we can construct an
action. If we denote the field strength associated with the generator
$G$ of $SO(1,5)$ by $\mathbb{F}_G$, then, using $\mathbb{F} =
d\mathbb{A} + \mathbb{A} \we \mathbb{A}$ we find,
  \eq{
  \mathbb{F}^{ab}_J = & \,\frac{1}{2} \( R^{ab} - \ell^{-2}\, e^a \we e^b - \ell^{-2}\, t^a \we t^b  \), \label{curvatureJ} \\
  \mathbb{F}^{a}_{P^{(1)}} = & \,\ell^{-1} \( \D_{\om} e^a  +  A \we t^a \),  \label{curvatureP} \\
  \mathbb{F}^{a}_{P^{(2)}} = & \,\ell^{-1} \( \D_{\om} t^a  -  A \we e^a \), \label{curvatureK}  \\
  \mathbb{F}_D = & \,d A + \ell^{-2} t^a \we e_a, \label{curvatureA}
  }
where $R^{ab} = d \om^{ab} + \om^a{}_c \we\om^{cb}$ is the Riemann
curvature. In particular, note that the field strengths associated with the
$P^{(i)}$ generators are generalizations of torsion associated to each
of the vielbeins. The infinitesimal transformation of the curvatures
under the action of the $SO(1,5)$ group mimic the transformation of
the gauge fields, namely, 
  \eq{
  \d_{\l} \mathbb{F}^{ab}_J =& - 2\ll_c{}^{[a} \mathbb{F}^{b]c}_J +  \chi^{[a} \mathbb{F}^{b]}_{P^{(1)}} +  \z^{[a} \mathbb{F}^{b]}_{P^{(2)}} \\
   \d_{\l} \mathbb{F}^{a}_{P^{(1)}} =& -\ll^{a}{}_b \mathbb{F}^b_{P^{(1)}} - \xi\, \mathbb{F}^a_{P^{(2)}} + 2 \chi_b \mathbb{F}^{ab}_J + \z^a \mathbb{F}_D ,  \\
   \d_{\l} \mathbb{F}^{a}_{P^{(2)}} =& -\ll^{a}{}_b \mathbb{F}^b_{P^{(2)}}  + \xi\, \mathbb{F}^a_{P^{(1)}} + 2 \z_b \mathbb{F}^{ab}_J - \chi^a \mathbb{F}_D,  \\
   \d_{\l} \mathbb{F}_D =& - \z_a \mathbb{F}^a_{P^{(1)}} +\chi_a \mathbb{F}^a_{P^{(2)}} ,
  }
except that now all quantities transform homogeneously under the
Lorentz and $SO(2)$ subgroups of $SO(1,5)$. In fact, both
$\mathbb{F}^{ab}_J$ and $\mathbb{F}_D$ are left invariant under
$SO(2)$ transformations. 


\subsection{Action and constraints}

We now construct an action involving the $SO(1,5)$ curvatures given in
eqs.~\eqref{curvatureJ} --~\eqref{curvatureA} that preserves parity and is invariant under the local $SO(1,3) \times SO(2)$ transformations singled out in the previous section. This is
analogous to the $SO(2,4)$ formulation of conformal
gravity but our construction differs from the
standard approach of~\cite{Kaku:1977pa}:  while the spin
connection $\omega^{ab}$ was assumed to be real in the previous section, from now on we will regard it as a complex quantity, $\omega^{ab}=\tau^{ab}+i\sigma^{ab}$. As a consequence the Riemann curvature $R^{ab}$ and the field strength given in eq.~\eqref{curvatureJ} are now complex. The
justification for this, and the meaning of the complex connection will
be explained in what follows. Hence, with some hindsight we consider
the action,
  \eq{
  I = M_p^2 \ell^2 \int \mathrm{Re} \( \mathbb{F}^{ab}_J \we \mathbb{F}^{cd}_J\) \e_{abcd} - M_p^2 \ell^2 \,\frac{\s^2}{2} \int \mathbb{F}_D \we \star\, \mathbb{F}_D, \label{gaugeaction}
  }
where $\e_{abcd}$ is the totally antisymmetric tensor, $\s^2$ is a
dimensionless parameter greater than one, and the Hodge dual is
defined with respect to the $SO(2)$-invariant metric $G_{\mu\nu}$
given in eq.~\eqref{so2metric}. In this action we have not used the
curvatures associated with the $P^{(i)}$ generators of
$SO(1,5)$. Instead, these will be used to impose constraints on the
spin connection that determine it in terms of the vielbeins.  

The first term in eq.~\eqref{gaugeaction} is also featured in the
gauge theory constructions of Einstein and conformal gravity of
refs.~\cite{MacDowell:1977jt,Kaku:1977pa} with the exception that
there $\mathbb{F}^{ab}_J$ is strictly real. This is the \emph{only}
term involving any of the curvatures that satisfies the above
conditions on symmetry and parity, and which does not require a
metric. On the other hand, the second term in eq.~\eqref{gaugeaction}
is a non-gravitational action which couples to the $SO(2)$-invariant
metric. This term must accompany the gravitational action in order to
render the theory free of ghosts to linear order around a de Sitter
background.

In the action~\eqref{gaugeaction} we have assumed that the spin
connection is \emph{complex} and, furthermore, that it is subject to
a \emph{constraint}. Let us explain why this must be the case. If we
assume that the spin connection is real, then its equations of motion
read, 
  \eq{
 \( e^a \we \mathbb{F}^b_{P^{(1)}} +
 t^a \we \mathbb{F}^b_{P^{(2)}} \) \e_{abcd} = 0. \label{omegaeom} 
  }
This equation is invariant under $SO(2)$ transformations, a
property that will carry over to the spin connection, in agreement
with eq.~\eqref{deltaomega}. Note, however, that the vector field $A$
does not appear in the above equation and is not a part of the
gravitational action. Then the latter will contain at most two spin-2 fields and
cannot have PM interactions beyond cubic order~\cite{Zinoviev:2006im,deRham:2013wv,Apolo:2016vkn}. To circumvent the
obstructions to non-linear partially massless symmetry we expect the
vector field to play a non-trivial role in the theory. Furthermore,
while it is difficult to solve for the spin connection in
eq.~\eqref{omegaeom}, we can readily find a perturbative solution
around a de Sitter background. We then find that, not only does the
vector field play no role in the gravitational action but, to linear
order in the fields, the latter propagates only a massless spin-2
field.\footnote{This means that the theory is strongly coupled on the
de Sitter background. In this calculation we assumed that the
vielbeins obey the symmetrization condition discussed below.}

Thus the gravitational action in eq.~\eqref{gaugeaction} is not to be
interpreted as a first order action, i.e.~one linear in derivatives,
but as a second order action where the spin connection obeys a
constraint. This is exactly what happens in the gauge theory approach
to conformal gravity, based on the $SO(2,4)$ group, where one
imposes~\cite{Kaku:1977pa},  
  \eq{
  \mathbb{F}^a_{K} = 0, \label{so24constraint}
  }
for some $SO(2,4)$ generator $K$ in the complement of $SO(1,3) \times
SO(1,1)$. It could be argued that this is also true for Einstein
gravity which is based on $SO(1,4)$~\cite{MacDowell:1977jt}. Indeed,
in this case there is only one generator $K$ in the complement of
$SO(1,3) \subset SO(1,4)$ for which eq.~\eqref{so24constraint} becomes the
torsionless condition which also coincides with the equations
of motion derived from the first order action. Note that in
contrast to the four dimensional case, the gauge theory approach to
Einstein and conformal gravity in three dimensions does not require
additional
constraints~\cite{Achucarro:1987vz,Witten:1988hc,Horne:1988jf}.

For the $SO(1,5)$ group, imposing the constraint in
eq.~\eqref{so24constraint} with $K $ given by any \emph{real} linear
combination of $P^{(1)}$ and $P^{(2)}$ does not lead to an
$SO(2)$-invariant spin connection. The reason being that the corresponding
curvatures, $\mathbb{F}^a_{P^{(1)}}$ and $\mathbb{F}^a_{P^{(2)}}$,
form and $SO(2)$ vector. While imposing $\mathbb{F}^a_{P^{(1)}}
= \mathbb{F}^a_{P^{(2)}} = 0$ is $SO(2)$ invariant, the only solution
to these equations is $\om^{ab} = 0$.

On the other hand, $SO(2)$-invariant  constraints can be easily
constructed in a basis of complex vielbeins  $\psi^a$ defined by,
  \eq{
  \psi^a = e^a + i  t^a. \label{complexvielbein}
  }
Indeed, a constraint that is invariant under $SO(2)$ transformations is given by,
  \eq{
  \mathbb{F}^a_{H} & \equiv  \mathbb{F}^a_{P^{(1)}} + i  \mathbb{F}^a_{P^{(2)}} = \ell^{-1} \( \D_{\om} - i A\) \we \psi^a = 0, \label{so15constraint}
  }
where $\mathbb{F}_H^a$ denotes the curvature associated to the complex vielbein~\eqref{complexvielbein}. We recognize in eq.~\eqref{so15constraint} the torsionless condition for a {\it complex} spin connection which is also invariant under the
local $U(1)$ transformations of $\psi^a$. Thus, the justification for
extending the $SO(1,5)$ action to a complex connection stems from the requirement of an $SO(2)$-invariant $\om^{ab}$ that satisfies non-trivial constraints. Note that the relevant formulas in section~\ref{fieldsANDcurvatures} can be easily extended to the
complex connection case and may be derived by complexifying the
Lorentz algebra in eqs.~\eqref{so15algebra1} --~\eqref{so15algebra4}, see Appendix~\ref{enhancedalgebra}.

If we assume invertibility of the complex vielbein, the spin
connection is then given by,  
  \eq{
    \om_{\mu}^{ab} = \psi^{\g a} D_{[\mu}\psi^b{}_{\g]} - \psi^{\g b} D_{[\mu}\psi^a{}_{\g]} - \psi^{\rho a} \psi^{\s b} \psi_{c\mu}{} D_{[\rho}\psi^c{}_{\s]}, \label{spinconnection}
  }
where $D_{\mu} = \p_{\mu} - i A_{\mu}$ is the covariant derivative
with respect to $A_{\mu}$, and $\psi^{\mu a}$ satisfies, 
  \eq{
  \psi^a{}_{\mu} \psi^{\mu}{}_b= \d^a_b, \qquad \psi^{\mu}{}_a \psi^a{}_{\nu} = \d^{\mu}_{\nu}.
  }
In particular, note that the spin connection in eq.~\eqref{spinconnection}
reduces to that of General Relativity when we turn off the vector
field and decouple either one of the vielbeins. 

There is an additional constraint we will impose on the vielbeins. In the
vielbein formulation of General Relativity the local Lorentz symmetry
can be used to remove all antisymmetric components from the
vielbein, insuring that no antisymmetric rank-2 tensors propagate
in the metric formulation of the theory. However, once an additional
vielbein is added, the local Lorentz symmetry can be used to remove
only one of the antisymmetric rank-2 tensors. Put in a different way,
bimetric theories in the vielbein and metric formulations are not
equivalent unless we impose the following symmetrization
constraint~\cite{Zumino:1970tu,Hinterbichler:2012cn} (see
also~\cite{Deffayet:2012zc}),\footnote{Physically, this condition
insures that the null cones of the two vielbeins always
intersect, allowing for a consistent spacetime
decomposition~\cite{Hassan:2016abc}.}  
  \eq{
  \(\begin{array}{cc} e^a{}_{\mu} &  t^a{}_{\mu} \end{array}\) 
\(\begin{array}{cc} 0 & -1 \\ 1 & 0 \end{array}\)\(\begin{array}{c}
e_{a\nu} \\  t_{a\nu} \end{array}\) = 0. \label{vielbeinconstraint}
  }
As a result of this, the vielbein contribution to the $\mathbb{F}_D$
term in the action \eqref{gaugeaction} vanishes. We have written the
constraint in a manifestly $SO(2)$-invariant way to highlight that it
is compatible with the symmetries of the action. Another 
consequence of this constraint is that the inverse $\psi^\mu{}_a$ of the complex vielbein
$\psi^a{}_\mu$ can be easily evaluated as,   
  \eq{
\psi^\mu{}_a=G^{\mu\nu}\psi^*_{a\nu},
  }
where $G^{\mu\nu}$ is the inverse of the $SO(2)$-invariant metric
  \eqref{so2metric}.
\\

\emph{To summarize}, we have constructed a gauge theory based on
$SO(1,5)$ that is invariant under local $SO(1,3) \times SO(2)$
transformations. The basic fields of this theory are a complex spin
connection $\om^{ab}$, a gauge field $A$, and two vielbeins $e^a$ and
$t^a$ from which we can construct charged~\eqref{fgmetrics} and
singlet~\eqref{so2metric} metrics.  The theory is described by the
action given in eq.~\eqref{gaugeaction} with constraints on the spin
connection~\eqref{so15constraint} and the
vielbeins~\eqref{vielbeinconstraint}.


\subsection{Geometric interpretation}

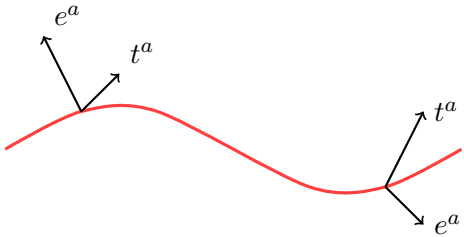
\begin{figure}[!h]
  \begin{center}
  \begin{tikzpicture}[scale=2]
  \draw [red!75,very thick] plot [smooth, thick, tension=.5, line width = 10mm] coordinates {(.5,.25) (1,.5) (1.5,.5) (2.5,0) (3,0)   (3.5,.25)};
  \draw [black, thick,->]  (1,.5) -- (.75,1) node[anchor=south west]{$e^a$};
  \draw [black, thick, ->]  (1,.5) -- (1.25,.75) node[anchor=south west,black]{$t^a$};
  \draw [black, thick, ->]  (3,0) -- (3.25,.5) node[anchor=west]{$t^a$};
  \draw [black, thick, ->]  (3,0) -- (3.25,-.25) node[anchor=west]{$e^a$};
  \end{tikzpicture}
  \caption{In the $SO(1,5)$ theory the $e^a$ and $t^a$ vielbeins mix linearly under parallel transport.}
  \label{theoneandonlyfigure}
  \end{center}
  \end{figure}

Let us now discuss the geometrical interpretation of the
complexification of the spin connection. We first note that
eq.~\eqref{so15constraint} defines how the covariant derivative acts
on the complex vielbein, i.e.,
  \eq{
  \D \psi^a \equiv \D_{\om} \psi^a = d \psi^a + \om^{a}{}_b \we \psi^b. \label{complexD}
  }
If we let $\om^{ab} = \tau^{ab} + i \s^{ab}$ where $\tau^{ab}$ and
$\s^{ab}$ are two real one-forms antisymmetric in $a$ and $b$, then
eq.~\eqref{complexD} implies that, 
  \eq{
  \D \(\begin{array}{c} e^a \\  t^a \end{array}\) \equiv  \(\begin{array}{cc} d & 0 \\ 0 & d \end{array}\)\(\begin{array}{c} e^a \\  t^a \end{array}\) + \(\begin{array}{cc} \tau^{a}{}_b & -\s^{a}{}_b \\ \s^{a}{}_b & \tau^{a}{}_b \end{array}\) \we \(\begin{array}{c} e^b \\  t^b \end{array}\). \label{realD}
  }
Thus, the complexification of the spin connection introduces an
additional real connection $\s^{ab}$ that is responsible for mixing
the $e^a$ and $t^a$ vielbeins. In particular, this mixing guarantees
that the covariant derivative of the vielbeins transforms
homogeneously under \emph{global} $SO(2)$ transformations. To
guarantee that the covariant derivative transforms homogeneously under
local $SO(2)$ transformations we must also add the vector field in the
obvious way, i.e. via a skew-symmetric matrix. The addition of this
spin connection is what makes the $SO(1,5)$ theory a bimetric theory
of gravity. Indeed, the bimetric theories of ref.~\cite{Hassan:2011zd}
also feature two spin connections which, in contradistinction to the
$SO(1,5)$ theory, act diagonally on the corresponding
vielbeins~\cite{Hinterbichler:2012cn}, 
  \eq{
  \D \(\begin{array}{c} e^a \\  t^a \end{array}\) \equiv  \(\begin{array}{cc} d & 0 \\ 0 & d \end{array}\)\(\begin{array}{c} e^a \\  t^a \end{array}\) + \(\begin{array}{cc} \tau^{a}{}_b & 0 \\ 0 & \s^{a}{}_b \end{array}\) \we \(\begin{array}{c} e^b \\  t^b \end{array}\). \label{vielbeintransformation2}
  }
This property of the covariant derivative is consistent with the fact
that the $e^a$ and $t^a$ vielbeins in the bimetric theory
of ref.~\cite{Hassan:2011zd} rotate independently of each other under
parallel transport. In contrast, in the $SO(1,5)$ theory the
$e^a$ and $t^a$ vielbeins mix linearly with each other as implied by
eq.~\eqref{realD} and illustrated in fig.~\ref{theoneandonlyfigure}
(this is over and above the simple $SO(2)$ rotation generated by the
gauge field $A$).
\\

It is important to note that while we do complexify the spin
connection in eq.~\eqref{gaugefield}, we \emph{do not} complexify the
Lorentz transformations parametrized by $\ll^{ab}$ in
eq.~\eqref{gaugetransformation}.\footnote{The connection $\s^{ab}$ may be associated with additional gauge transformations not considered here.} In particular, eq.~\eqref{deltaomega} implies
that, under local Lorentz transformations, the real spin connections
transform as,
  \eq{
  \d_{\ll} \tau^{ab} &= \D_{\tau} \ll^{ab},  \qquad \d_{\ll} \s^{ab} = \s^{a}{}_{c} \ll^{cb} + \s^{b}{}_{c} \ll^{a c}. \label{deltaomega2}
  }
Thus, the $\s^{ab}$ connection transforms homogeneously, i.e.~as a
tensor, under local Lorentz transformations. Eq.~\eqref{deltaomega2}
guarantees that the covariant derivative defined in
eq.~\eqref{vielbeintransformation} transforms homogeneously as
well. In particular, these transformation properties of the
connections guarantee that the real part of the Riemann curvature,
which enters the action defined in eq.~\eqref{gaugeaction}, also
transforms covariantly under local $SO(1,3)$ transformations, 
  \eq{
  \mathrm{Re}\, R^{ab}(\omega) = R^{ab}( \tau) - \s^{a}{}_c \we \s^{c b}, \qquad \d_{\ll} \mathrm{Re} \,R^{ab}(\omega)=-2 \ll_c{}^{[a} \mathrm{Re}\, R^{b] c}(\omega). \label{realRw}
  }
Here $R^{ab}( \tau)=d\tau^{ab}+\tau^a_{~c}\wedge\tau^{cb}$ is the
Riemann curvature for the real part of the spin connection. Thus, the
theory can be described either in terms of a complex connection, or a
real connection on the bundle $e\oplus t$. This is related to the complexification of the Lorentz algebra as discussed in more detail in Appendix~\ref{enhancedalgebra}.  
\\

\noindent{\bf Analytic continuation of conformal gravity.}
We conclude this section by discussing the relationship between the
$SO(1,5)$ theory we have constructed and the $SO(2,4)$ theory of
ref.~\cite{Kaku:1977pa}. Let us first note that the generator $H$
associated with the constraint in eq.~\eqref{so15constraint} is given
by, 
  \eq{
  H_a = P^{(1)}_a - i P^{(2)}_a.
 }
This is part of the analytic continuation that turns the $\mathfrak{so}(1,5)$
algebra described in eqs.~\eqref{so15algebra1} --~\eqref{so15algebra4}
into $\mathfrak{so}(2,4)$. Indeed, if we let 
  \eq{
  \tilde{P}^{(1)}_a = H_a, \qquad \tilde{P}^{(2)}_a = H^{*}_a, \qquad \tilde{D} = i D,
  }
we recover the algebra of the conformal group which is used in the
construction of conformal gravity~\cite{Kaku:1977pa}. In particular, the
constraint~\eqref{so15constraint} corresponds to the analytic
continuation of the constraint~\eqref{so24constraint} imposed
in~\cite{Kaku:1977pa}, modulo conventions. This observation also
extends to the gravitational part of the action given in
eq.~\eqref{gaugeaction}. Using eq.~\eqref{curvatureJ} and the complex
vielbein~\eqref{complexvielbein}, the latter may be written as,
  \eq{
  I_G = -\frac{M_p^2}{2} \int \e_{abcd} \bigg \{ \psi^a \we \psi^{* b} \we \re\, R^{cd} - \frac{1}{2\ell^2} \psi^a \we \psi^{* b} \we \psi^c \we \psi^{* d} \bigg \},
  }
where the higher derivative term $\int
R^{ab} \we R^{cd}\e_{abcd}$ is topological and does not contribute to
the action. This is the same action obtained in~\cite{Kaku:1977pa} where,
instead, $\psi^a$ and $\psi^{* a}$ are treated as two independent real vielbein one-forms, and the real spin connection obeys
eq.~\eqref{so15constraint} with $A \ra - i \tilde{A}$. Thus, the \emph{extended} $SO(1,5)$ theory considered here corresponds to an analytic
continuation of the $SO(2,4)$ theory that leads to conformal gravity. Note
that while it is possible to integrate out, say, $\psi^{* a}$ in the
$SO(2,4)$ theory, and thereby obtain conformal gravity, this is no longer
the case in the $SO(1,5)$ theory since $\psi^{a}$ and $\psi^{* a}$ are
complex conjugates of each other. Furthermore, unlike conformal gravity,
the $SO(1,5)$ theory contains an additional vector field.


\subsection{Bimetric formulation}

Let us now consider the action~\eqref{gaugeaction} in more
detail. Using eqs.~\eqref{curvatureJ},~\eqref{curvatureA}, and the
constraint~\eqref{vielbeinconstraint}, the action can be written concisely in
terms of complex vielbeins,  
  \eq{
  I = -\frac{M_p^2}{2} \int \e_{abcd} \bigg \{ \psi^a \we \psi^{* b} \we \Big ( \re\, R^{cd} - \frac{1}{2\ell^2}  \psi^c \we \psi^{* d} \Big) \bigg \} - M_p^2 \ell^2\,\frac{\s^2}{2} \int F \we \star\, F,  \label{vielbeinaction}
  }
where $F = dA$ is the field strength of the vector field. This action
describes a bimetric theory in the vielbein formulation with
non-standard kinetic terms but the same potential as the bimetric
models studied in 
refs.~\cite{Hassan:2012gz,Hassan:2012rq,Hassan:2013pca,Hassan:2015tba} in connection with PM symmetry.  

To see this more explicitly, let us express the action in 
\eqref{vielbeinaction} using metric variables. It is easy to
verify that when $e_{[\mu}{}^a t_{\nu] a}=0$, as implied by the
symmetrization constraint~\eqref{vielbeinconstraint}, one can write, 
\eq{
  S^{\mu}{}_{\nu} =  \big(\sqrt{g^{-1}f} \,\big)^{\mu}{}_{\nu}
  =e^\mu{}_{a} t^a{}_{\nu}.
  }
Then the action~\eqref{vielbeinaction} can be written as (see ref.~\cite{Hinterbichler:2012cn} for details),
  \eqsp{
  I = & \,M_p^2 \int d^4x \absq{g}\, e^{\mu}{}_a e^{\nu}{}_b\,\re\, R^{ab}{}_{\mu\nu} + M_p^2 \int d^4x \absq{f}  \,t^{\mu}{}_a t^{\nu}{}_b \, \re\,R^{ab}{}_{\mu\nu} \\
  & - 2 M_p^2 \ell^{-2}\int d^4x \absq{g}\, \sum_{n=0}^{4} \b_n e_n (S) -  M_p^2 \ell^2\,\frac{\s^2}{2} \int d^4x \absq{G} G^{\mu\a} G^{\nu\b}\, F_{\mu\nu} F_{\a\b}, \label{metricaction}
  }
where the $g_{\mu\nu}$, $f_{\mu\nu}$, and $G_{\mu\nu}$ metrics are
defined in eqs.~\eqref{fgmetrics} and~\eqref{so2metric}. In eq.~\eqref{metricaction} the $\b_n$
  parameters are given by
  \eq{
  \b_0 = 3 , \qquad \b_2 = 1, \qquad \b_4 = 3, \qquad \b_1 = \b_3 = 0, \label{PMbetaparameters}
  }
and $e_n(S)$ denote the elementary symmetric polynomials of the
  matrix $S$. In particular, 
  \eq{
  e_0(S) &  = 1, \qquad    e_2(S) = \frac{1}{2} \Big [ \tr(S)^2
  - \tr(S^2) \Big ], \qquad e_4(S) =\det S= \frac{\absq{f}}{\absq{g}}.
  }

The kinetic terms in eq.~\eqref{metricaction} are different from those
of bimetric gravity which are given by the Einstein-Hilbert action for
each of the metrics. Furthermore, by rescaling $t^a \ra \a t^a$ and
taking the limits $\s \ra \infty$ and either $\a \ra 0$ or
$\a \ra \infty$ with $\ell_1^{-2} \a^2$ finite, we can decouple either
one of the metrics in eq.~\eqref{metricaction} and recover the
Einstein-Hilbert action with a positive cosmological
constant. In particular, note that once we decouple the vector field, the
limits $\a \ra 0$ and $\a \ra \infty$ make the spin connection real
and reduce it to that of General Relativity for either one of the
vielbeins.

On the other hand, the potential in eq.~\eqref{metricaction} is the
same as that of the ghost-free bimetric model of
refs.~\cite{Hassan:2012gz,Hassan:2012rq,Hassan:2013pca,Hassan:2015tba},
investigated in connection with PM symmetry.  This model is singled
out among all ghost-free bimetric gravity actions by demanding
invariance under a global version of the partially massless symmetry
-- one given by eq.~\eqref{pmsymmetry} with constant $\xi$ -- beyond
the linear level. This condition leads to eq.~\eqref{PMbetaparameters}
which uniquely fixes the $\b_n$ parameters, up to constant scalings of
the metrics.  However, as shown in~\cite{Apolo:2016vkn}, when $\xi$ is
a function of the coordinates, the partially massless symmetry cannot
be extended beyond cubic order in the would-be partially massless
field.

One can now understand how the $SO(1,5)$ theory avoids this
obstruction. In bimetric gravity, the potential satisfying
eq.~\eqref{PMbetaparameters} is locally $SO(2)$-invariant in the
vielbein formulation.\footnote{The $SO(2)$
invariance of the bimetric potential, and its possible
significance to PM symmetry, was first noticed by Latham
Boyle and investigated with Kurt Hinterbichler and Angnis Schmidt-May~\cite{Boyle:2015abc}.} However, the kinetic terms explicitly
break this local symmetry of the potential. In the present
construction, the constraint \eqref{so15constraint} introduces an
additional vector field that allows us to construct locally $SO(2)$-invariant kinetic terms. The presence of the vector field drastically
modifies the structure of the cubic and higher order interactions,
thus circumventing the known obstructions to non-linear PM symmetry. 
As will be shown later, the PM symmetry is present to all orders and
is simply a transmutation of the local $SO(2)$ symmetry of the
$SO(1,5)$ theory. 

Let us conclude this section with comments on the possible ghost
instabilities of the theory. While the potential
in~\eqref{metricaction} does not introduce ghost instabilities in the
standard bimetric gravity~\cite{Hassan:2011ea}, this has only been
proved when the kinetic terms of the latter are described by the
Einstein-Hilbert action for each of the metrics. Thus, the
non-standard kinetic terms of the $SO(1,5)$ theory have the potential
to introduce ghosts, unless additional constraints exist that remove
these modes. Below we will show that the theory is indeed ghost free
at linear order. Furthermore, note that similar kinetic terms have  
been considered recently in the
literature~\cite{Folkerts:2011ev,Hinterbichler:2013eza,deRham:2015rxa,Akhavan:2016xmj} 
and these have been shown to contain nonlinear constraints that remove 
propagating modes also beyond linear
order~\cite{Li:2015izu,Li:2015iwc}. This opens up the possibility that
the $SO(1,5)$ theory remains ghost free to all orders.


Finally, note that the coupling to matter fields via the
$SO(2)$-invariant metric $G_{\mu\nu}$, or even via the charged
metrics, has the potential to introduce ghost instabilities as
well. We will see that perturbatively around a de Sitter background,
the $SO(2)$-invariant metric contains only the massless spin-2 mode
and does not induce couplings between matter and the massive spin-2
field. While coupling matter to the massless mode is not ghost free in
bimetric gravity~\cite{Hassan:2012wr}, the appearance of ghosts at the
non-linear level depends on the structure of the kinetic terms, which
are different in the $SO(1,5)$ theory and not investigated here.


\section{Perturbative metric formulation}
\label{se:pmsymmetry}

In this section we expand the action around an off-shell,
$SO(2)$-invariant metric and express the result in terms of
spin-2 variables. We find that to linear order around a de Sitter
background the theory propagates a massless spin-2 field, a partially
massless graviton, and a massless vector field. We then show that the
partially massless symmetry found at linear order in the metric
formulation is nothing but the local $SO(2)$ symmetry manifest to all
orders in the vielbein formulation of the theory.


\subsection{Background solutions}

We begin by considering the background solutions to the equations of
motion. First, it is important to note that imposing the
symmetrization constraint~\eqref{vielbeinconstraint} before and after
varying the action can lead to different equations of motion. The
reason is that terms linear in the constraint give contributions to
the equations of motion that are not proportional to the constraint
itself. In the $SO(1,5)$ theory we must impose the symmetrization
constraint directly in the action. However, this is not an easy task
due to the non-linear structure of the kinetic term which mixes the
$e^a$ and $t^a$ vielbeins. It is nevertheless possible to derive the
equations of motion consistently provided we complement the variation
of the action by appropriate ``counterterms''. Thus, in order to take
care of possible terms linear in the constraint, we determine the
equations of motion from the following variation of the action,
  \eq{
  \d \tilde{I} = \d I - \int d^4 x \frac{\d I}{\d Q^{\mu\nu}} \d Q_{\mu\nu}, \label{modvar}
  }
where $I$ is the action of the $SO(1,5)$ theory given in eq.~\eqref{vielbeinaction} and $Q_{\mu\nu} = e_{a[\mu} t^a{}_{\nu]}$ is the symmetrization constraint given in eq.~\eqref{vielbeinconstraint}. The modified variation of the action~\eqref{modvar} allows us to impose $Q_{\mu\nu} = 0$ either before or after we obtain the equations of motion.

The variation of the action is then given by,
  \eq{
  \d \tilde{I} = - M_p^2 \int d^4x \( X^{\mu}{}_a \d e^a{}_{\mu} + Y^{\mu}{}_a \d t^a{}_{\mu} + Z^{\mu} \d A_{\mu} + \frac{1}{M_p^2} \frac{\d I}{\d Q^{\mu\nu}} \d Q_{\mu\nu} \),
  }
where the functions $X^{\mu}{}_a, Y^{\mu}{}_a$, and $Z^{\mu}$ read,
  \begin{align}
  \begin{split}
  	X^{\mu}{}_a = \e_{ijcd} \, \e^{\l\nu\a\b} \bigg \{ & \d^i_a \d^{\mu}_{\l} e^j{}_{\nu} \Big [ \frac{1}{2} \re\,R^{cd}{}_{\a\b}  -  \ell^{-2} \big ( e^c{}_{\a} e^d{}_{\b} + t^c{}_{\a} t^d{}_{\b}  \big )  \Big ] \\
  + & \frac{1}{4}\, \( \frac{\d \om^{ij}_{\l}}{\d \psi^a{}_{\mu}} \D_{\nu} + \frac{\d \om^{* ij}_{\l}}{\d \psi^{*a}{}_{\mu}} \D^{*}_{\nu} \) \( e^c{}_{\a} e^d{}_{\b} + t^c{}_{\a} t^d{}_{\b}\)\bigg \} - \frac{1}{M_p^2} \frac{ \d I_A}{ \d e^a{}_{\mu}}, \label{eeom}
  	\end{split}\\
		\begin{split}
  Y^{\mu}{}_a = \e_{ijcd} \, \e^{\l\nu\a\b} \bigg \{ & \d^i_a \d^{\mu}_{\l} t^j{}_{\nu} \Big [ \frac{1}{2}\, \re\,  R^{cd}{}_{\a\b} - \ell^{-2} \big (  t^c{}_{\a} t^d{}_{\b} + e^c{}_{\a} e^d{}_{\a}  \big )  \Big ] \\
  + & \frac{i}{4}\( \frac{\d \om^{ij}_{\l}}{\d \psi^a{}_{\mu}} \D_{\nu} - \frac{\d \om^{* ij}_{\l}}{\d \psi^{*a}{}_{\mu}} \D^{*}_{\nu} \) \( e^c{}_{\a} e^d{}_{\b} + t^c{}_{\a} t^d{}_{\b} \) \bigg \}  - \frac{1}{M_p^2} \frac{ \d I_A}{ \d t^a{}_{\mu}},  \label{teom}
  \end{split}\\
  Z^{\mu} = \e_{ijcd} \, \e^{\l\nu\a\b} i & \Big ( \psi^{\mu [i}  \psi^{j]}{}_{\l}\, \D_{\nu} - \psi^{* \mu [i}  \psi^{*j]}{}_{\l}\, \D_{\nu}^{*} \Big ) \(e^c{}_{\a} e^d{}_{\b} + t^c{}_{\a} t^d{}_{\b} \) - \frac{2}{M_p^2} \frac{\d I_A}{\d A_{\mu}}.
  \end{align}
In these equations $I_A = - M_p^2 \ell^2\,\frac{\s^2}{2} \int F \we \star F$, whose variations are proportional to $A$ and will not be needed in what follows. Furthermore, in eqs.~\eqref{eeom} and~\eqref{teom} $\d \om^{ab}_{\mu}/ \d \psi^c_{\l}$ is an operator that contains $U(1)$ covariant derivatives $D_{\mu} = \p_{\mu} - i A_{\mu}$ acting on terms to its right, and $\D_{\mu}$ is the covariant derivative with respect to the complex spin connection.   

In close analogy to bimetric gravity~\cite{Hassan:2012gz}, the $SO(1,5)$ theory admits proportional background solutions where
  \eq{
  e^a = \bar{e}^a, \qquad t^a = c\, \bar{e}^a, \qquad A = 0, \label{backgroundsol}
  }
and $c$ is an arbitrary constant. For this ansatz it is not necessary to evaluate $\d I / \d Q^{\mu\nu}$ explicitly and the equations of motion become,
  \eq{
  X^{\mu}{}_a = 0, \qquad Y^{\mu}{}_a = 0, \qquad Z^{\mu} = 0.
  }
Furthermore, for the parametrization given in eq.~\eqref{backgroundsol} the spin connection is real and satisfies the torsionless condition with respect to the background vielbein $\bar{e}^a$,
  \eq{
  \D \bar{e}^a = d \bar{e}^a + \om^{a}{}_b \we \bar{e}^b = 0 \quad \Rightarrow \quad \om_{\mu}^{ab} = \bar{e}^{\g a} \p_{[\mu}\bar{e}^b{}_{\g]} - \bar{e}^{\g b} \p_{[\mu}\bar{e}^a{}_{\g]} - \bar{e}^{\rho a} \bar{e}^{\s b} \bar{e}_{c \mu} \p_{[\rho} \bar{e}^c{}_{\s]}. \label{backgroundspinconnection}
  }
This implies that the second line in each of eqs.~\eqref{eeom} and~\eqref{teom} vanish, while the $Z^{\mu}$ equation of motion is satisfied exactly. The remaining equations of motion are equal to each other and reduce to,
  \eq{
   0 & = \e_{abcd} \e^{\mu\nu\a\b} \, \bar{e}^b {}_{\nu} \Big [ R^{cd}{}_{\a\b} - 2\ell^{-2} \(1+ c^2\) \,\bar{e}^c{}_{\a} \bar{e}^d{}_{\b} \Big ], \label{backgroundeom}
  }
where $R^{ab}{}_{\a\b}$ is the Riemann curvature for the real spin connection given in eq.~\eqref{backgroundspinconnection}. Eq.~\eqref{backgroundeom} is equivalent to Einstein's equation $0 = R_{\mu\nu} - \frac{1}{2} R\, \bar{g}_{\mu\nu} + \ll\, \bar{g}_{\mu\nu}$ where $\bar{g}_{\mu\nu} = \bar{e}^a{}_{\mu}\bar{e}_{a \nu}$ and the cosmological constant is given by,
  \eq{
  \ll = 3\, \ell^{-2} \( 1+ c^2 \). \label{cosmologicalconstant}
  }
Thus, the $SO(1,5)$ theory admits asymptotically de Sitter backgrounds.
  
\subsection{Expansion of the action}

To identify the physical content of the theory we can now expand the
action perturbatively around a de Sitter background.  However, before
doing so, it is convenient to first obtain an expansion around a generic
off-shell $SO(2)$-invariant configurations. 

Besides the vector field $A$, our initial variables 
are the two vielbeins $e^a$ and $t^a$,
subject to the symmetrization constraint given in
eq.~\eqref{vielbeinconstraint}. From these vielbeins it is possible to
define two intermediate metric variables that transform non-trivially
under $SO(2)$ -- these are the $g_{\mu\nu}$ and $f_{\mu\nu}$ metrics
given in eq.~\eqref{fgmetrics}. The only $SO(2)$-invariant quantity is
the $G_{\mu\nu}$ metric given in eq.~\eqref{so2metric}. For
convenience we introduce the following rescaled version of the
$SO(2)$-invariant metric,
 \eq{
  G_{\mu\nu} = g_{\mu\nu} + f_{\mu\nu} = \(1 + c^2 \) \tilde{g}_{\mu\nu}.
  }
The metrics $g_{\mu\nu}$ and $f_{\mu\nu}$ can now be parametrized in
terms of a final, physical set of metric variables, namely 
\eq{
g_{\mu\nu}=\tilde g_{\mu\nu} -\phi_{\mu\nu} \,,\qquad f_{\mu\nu}=c^2\tilde g_{\mu\nu} +\phi_{\mu\nu} \label{metricexpansion}.
}
Here $\tilde{g}_{\mu\nu}$ is the spacetime metric whose perturbations
around a de Sitter background describe a massless spin-2 field. On the
other hand, by expanding the action in powers of $\phi_{\mu\nu}$ and
diagonalizing the quadratic terms, we will see that $\phi_{\mu\nu}$ is
a massive spin-2 field. Also note that up to a normalization
eq.~\eqref{metricexpansion} is the same expansion used in the bimetric
theory of
refs.~\cite{Hassan:2012gz,Hassan:2012rq,Hassan:2013pca,Hassan:2015tba}.

In order to write the action~\eqref{vielbeinaction} using metric
variables we need an explicit parametrization of the vielbeins in
terms of $\tilde{g}_{\mu\nu}$ and $\phi_{\mu\nu}$. We use
  \eq{
  e^a = \tilde{e}^a + \d e^a, \qquad t^a = c \( \tilde{e}^a + \d t^a \), 
  \label{smartexpansion}
  }
where $\tilde{e}^a$ is the $SO(2)$-invariant vielbein associated to
the $SO(2)$-invariant metric via 
  \eq{
  \tilde{g}_{\mu\nu} = \tilde{e}^a{}_{\mu} \,\tilde{e}_{a\nu}.
  }
In particular, if we let $\d e^a \ra 0$, $\d t^a \ra 0$, and $A \ra
0$, then $\tilde{g}_{\mu\nu}$ obeys Einstein's equations with a
positive cosmological constant, as shown in the previous section.
Note that the $SO(2)$ transformations of $\d e^a$ and $\d t^a$ in
eq.~\eqref{smartexpansion} can be worked out from
eq.~\eqref{vielbeintransformation} provided that we treat
$\tilde{e}^a$ as an $SO(2)$ singlet. It is also obvious that these
fields are not independent of each other and, in fact, are non-linear
in  $\phi_{\mu\nu}$. Indeed, from eqs.~\eqref{smartexpansion}
and~\eqref{metricexpansion} we find that,
  \eqsp{
\phi_{\mu\nu}&=-\left(\tilde{e}^a{}_{\mu} \d e_{a\nu}+\tilde{e}^a{}_{\nu}  
\d e_{a\mu} + \d e^a{}_{\mu} \d e_{a\nu} \right) =c^2\left(\tilde{e}^a{}_{\mu} \d t_{a\nu}+
\tilde{e}^a{}_{\nu} \d t_{a\mu} + \d t^a{}_{\mu} \d t_{a\nu }\right).
\label{deltag}
  }
Thus, eq.~\eqref{deltag} provides a non-linear relation between $\d
e^a$ and $\d t^a$.  Another non-linear relationship between $\d e^a$
and $\d t^a$ follows from the symmetrization constraint on the
vielbeins~\eqref{vielbeinconstraint}, which is responsible
for removing an antisymmetric rank-2 tensor from the metric
formulation of the theory.

We now solve the symmetrization constraint~\eqref{vielbeinconstraint} and
eq.~\eqref{deltag} perturbatively in the field $\phi_{\mu\nu}$. First, the vielbein symmetrization constraint is solved to quadratic order by,
  \eqsp{
  \d e^a{}_{\mu} & = \frac{1}{2} \d E_{\mu\nu} \tilde{e}^{\nu a} - \frac{1}{8} \d E_{\mu}{}^{\a}\d T_{\a \nu} \tilde{e}^{\nu a} + \dots, \\
    \d t^a{}_{\mu} & = \frac{1}{2} \d T_{\mu\nu} \tilde{e}^{\nu a} - \frac{1}{8} \d T_{\mu}{}^{\a}\d E_{\a \nu} \tilde{e}^{\nu a} + \dots, \label{deltaE}
  }
where all spacetime indices are raised with the $\tilde{g}_{\mu\nu}$
metric, and the fields $\d E_{\mu\nu}$ and $\d T_{\mu\nu}$ are symmetric rank-2
tensors. In these equations we have used the local Lorentz symmetry of the theory to make $\d E_{\mu\nu}$ symmetric. Then, to satisfy the symmetrization constraint $\d T_{\mu\nu}$ must be symmetric as well. In terms of these variables the solution to eq.~\eqref{deltag} is,
  \eqsp{
  \d E_{\mu\nu} & = -\phi_{\mu\nu} - \( \frac{1+c^2}{4c^2} \) \phi_{\mu}{}^{\a} \phi_{\a \nu}  + \dots, \\
    \d T_{\mu\nu} & = \frac{1}{c^2} \bigg [ \phi_{\mu\nu} - \( \frac{1+c^2}{4c^2} \) \phi_{\mu}{}^{\a} \phi_{\a \nu} \bigg ] + \dots. \label{deltaEdeltag}
  }
Eqs.~\eqref{deltaE} and~\eqref{deltaEdeltag} allow us to express the vielbeins in terms of the physical metric variables order by order in $\phi_{\mu\nu}$.

Finally, to express the action entirely in terms of metric variables
and their perturbations we need to define an appropriate covariant
derivative compatible with the $\tilde{g}_{\mu\nu}$ metric. We then
recall that in the absence of perturbations, i.e.~for $e^a
= \tilde{e}^a$, $t^a = c \tilde{e}^a$, and $A = 0$, the
action~\eqref{vielbeinaction} reduces to the Einstein-Hilbert action
for $\tilde g_{\mu\nu}$. Furthermore, in this case $\omega^{ab}_\mu$ becomes the spin connection given in eq.~\eqref{backgroundspinconnection} for
the vielbein $\tilde e^a$. Thus,
it is natural to define the covariant derivative via the vielbein
postulate for $\tilde{e}^a$,
  \eq{
\nabla_{\mu} \tilde{e}^a{}_{\nu} = \p_{\mu}  \tilde{e}^a{}_{\nu} + \om^{a}_{\mu b} \tilde{e}^b{}_{\nu} - \gg^{\l}_{\mu\nu} \tilde{e}^a{}_{\l}=0  \,\,\, \Rightarrow \,\,\, \gg^{\l}_{\mu\nu} = \frac{1}{2} \tilde{g}^{\l \a} \( \p_{\mu} \tilde{g}_{\nu \a} + \p_{\nu} \tilde{g}_{\mu \a} - \p_{\a} \tilde{g}_{\mu\nu}  \), \label{christoffelconnection}
  }
where $\gg^{\l}_{\mu\nu}$ is the Christoffel connection for $\tilde{g}_{\mu\nu}$. In particular, note that eq.~\eqref{christoffelconnection} is consistent provided that $\tilde{e}^a$ is a singlet under $SO(2)$ transformations.

With these ingredients in place we can now write the vielbein
action~\eqref{vielbeinaction} in a perturbative expansion of the metric formulation of the theory. Up to quadratic order in $\phi_{\mu\nu}$ we find,
  \eq{
  I =& \( 1+  c^2 \) M_p^2  \int d^4 x \absq{\tilde{g}} \bigg \{ R - 2 \ll + \frac{1}{2 c^2} \L_{PM}(\phi) - \frac{2}{ c} \big ( \phi_{\mu\nu} - \tilde{g}_{\mu\nu} \phi \big ) \nabla^{\mu} A^{\nu} + 6 A_{\mu}A^{\mu}  \bigg \} \notag \\ 
  &- \frac{M_p^2 \ell^2}{2} \s^2 \int d^4 x \absq{\tilde{g}} F_{\mu\nu} F^{\mu\nu}, \label{perturbativeaction}
  }
where $R$ is the Ricci scalar of $\tilde{g}_{\mu\nu}$ and $\phi
= \tilde{g}^{\mu\nu} \phi_{\mu\nu}{}$. In eq.~\eqref{perturbativeaction}, $\L_{PM}(\phi)$ is the
Lagrangian of a partially massless field defined on an arbitrary
metric $\tilde{g}_{\mu\nu}$, 
  \eq{
  \L_{PM}(\phi) = &  -  \frac{1}{2} \nabla_{\rho}\phi_{\mu \nu} \nabla^{\rho}\phi^{\mu \nu} + \frac{1}{2} \nabla_{\rho}\phi \nabla^{\rho}\phi -  \nabla_{\rho}\phi \nabla_{\s}\phi^{\rho \s} + \nabla_{\rho}\phi_{\mu \nu} \nabla^{\nu}\phi^{\mu \rho} + \frac{2\ll}{3} \phi_{\mu\nu} \phi^{\mu\nu}  \notag \\ 
    & -  \frac{\ll}{6} \phi^2 + \({\cal G}^{\mu \nu} + \ll \tilde{g}^{\mu\nu} \)\Big (- \frac{1}{2} \tilde{g}_{\mu \nu} \phi_{\rho \s} \phi^{\rho \s} + \frac{1}{4} \tilde{g}_{\mu \nu} \phi^2 + 2 \phi^{\rho}{}_{\nu} \phi_{\mu \rho} -  \phi\,  \phi_{\mu \nu} \Big), \label{pmaction}
  }
where ${\cal G}_{\mu\nu} = R_{\mu\nu} - \frac{1}{2} \tilde{g}_{\mu\nu} R$ is
the Einstein tensor. In particular, in a de Sitter background this
Lagrangian reduces to the Fierz-Pauli Lagrangian~\cite{Fierz:1939ix}
describing a massive spin-2 field where the mass of the graviton
saturates the Higuchi bound~\cite{Higuchi:1986py},
cf.~eq.~\eqref{higuchi}.  

Let us note that eq.~\eqref{pmaction} is the same quadratic Lagrangian
recovered from bimetric gravity in~\cite{Apolo:2016vkn} and which had
been used in~\cite{Joung:2014aba} to analyze the global symmetries of
partially massless gravity. However, this similarity between the
bimetric and $SO(1,5)$ theories does not extend to higher
orders. Indeed, already at cubic order and ignoring contributions from
the vector field, we find that the action of the $SO(1,5)$ theory
disagrees with the candidate PM bimetric theory studied in
refs.~\cite{Joung:2014aba,Apolo:2016vkn}. This means that the
realization of partially massless gravity discussed in the next
section cannot be obtained simply by adding a vector field to bimetric
gravity. As stressed previously, it is not only the vector field who
plays an important role in the theory, but also the kinetic terms used
in~\eqref{vielbeinaction} which render the action $SO(2)$
invariant.
   

\subsection{Partially massless symmetry}

Let us now recover the sought-after partially massless symmetry by diagonalizing the quadratic action given in eq.~\eqref{perturbativeaction}. We begin by expanding the
metric $\tilde{g}_{\mu\nu}$ around a de Sitter background, 
  \eq{
  \tilde{g}_{\mu\nu} = \bar{g}_{\mu\nu} + h_{\mu\nu},
  }
and performing the following field redefinition
  \eq{
  \phi_{\mu\nu} = \vp_{\mu\nu} - 6 \frac{c}{ \ll} \overline{\nabla}_{(\mu} A_{\nu)}. \label{fieldredefinition}
  }
While it is possible to diagonalize the action away from the de Sitter
background, this requires an additional redefinition of the metric
that removes higher derivative terms induced by
eq.~\eqref{fieldredefinition} at higher orders.\footnote{We have
checked that such field redefinitions exists at least up to cubic
order terms in the action.} These field redefinitions leave our
results unchanged at quadratic order so we can safely ignore them. 

Using the field redefinition given in eq.~\eqref{fieldredefinition} the
action~\eqref{perturbativeaction} reduces to 
  \eq{
  I= & M_p^2 \int d^4x \absq{\bar{g}} \bigg \{ \( 1+ c^2 \)   \bigg [ \L_{FP}(h) + \frac{1}{2 c^2} \L_{PM}(\vp) \bigg ] -  \ell^2\,\frac{\s^2 - 1}{2}  F_{\mu\nu} F^{\mu\nu}  \bigg \}, \label{diagonalperturbativeaction}
  }
where we must choose $\s^2 > 1$ in order to keep the action ghost-free
to quadratic order in the fields. In
eq.~\eqref{diagonalperturbativeaction} $\L_{FP}(h)$ is the Fierz-Pauli
Lagrangian~\cite{Fierz:1939ix} for a massless spin-2 field, while $\L_{PM}(\vp)$ is given by eq.~\eqref{pmaction} restricted to
the de Sitter background. We thus find that, to linear order on a de
Sitter background, the $SO(1,5)$ theory propagates a massless spin-2
field, a partially massless graviton, and a massless vector field. In
particular, the partially massless symmetry of $\vp_{\mu\nu}$ is
nothing but the local $SO(2)$ symmetry of the vielbein
action~\eqref{vielbeinaction}. In order to see this let us consider
the behavior of $h_{\mu\nu}$, $A_{\mu}$, and $\vp_{\mu\nu}$ under
infinitesimal $SO(2)$ transformations. From
eqs.~\eqref{vielbeintransformation} and~\eqref{smartexpansion} we find
that perturbations of the vielbeins transform as 
  \eq{
  (\d e^{a})' = \d e^a-  c \,\xi \( \tilde{e}^a + \d t^a \), \qquad (\d t^{a})' = \d t^a + \frac{1}{ c} \xi \( \tilde{e}^a + \d e^a \).
  }
Then, using eq.~\eqref{deltag}, the $SO(2)$ transformations of the fields
appearing in the perturbative formulation of the theory prior to
diagonalization are given by 
  \eq{
  \d_{\xi} h_{\mu\nu} & = 0, \label{deltah} \\
  \d_{\xi} A_{\mu\phantom{\nu}} & = \p_{\mu} \xi, \label{deltaa2} \\
  \d_{\xi} \phi_{\mu\nu} & = 2c \xi\,\bar{g}_{\mu\nu} + 2c \xi\, h_{\mu\nu} + \frac{(1 - c^2)}{ c} \xi\, \phi_{\mu\nu} + \dots, \label{deltaphi}
  }
where we have ignored higher order corrections in
$\d_{\xi} \phi_{\mu\nu}$ that depend on both $\phi_{\mu\nu}$ and
$h_{\mu\nu}$. In contrast, the transformations of $h_{\mu\nu}$ and
$A_{\mu}$ given in eqs.~\eqref{deltah} and~\eqref{deltaa2} are
exact. That $\d_{\xi} h_{\mu\nu} = 0$ is valid to all orders
follows from the $SO(2)$ invariance of $\tilde{g}_{\mu\nu}$.

On the other hand, the fields that diagonalize the action
via~\eqref{fieldredefinition} transform as,    
  \eq{
  \d_{\xi} h_{\mu\nu} & = \O(\vp), \label{deltah2} \\
  \d_{\xi} A_{\mu\phantom{\nu}} & = \p_{\mu} \xi, \label{deltaa3} \\
  \d_{\xi} \vp_{\mu\nu} & = \frac{6 c}{ \ll} \( \overline{\nabla}_{\mu} \overline{\nabla}_{\nu} + \frac{\ll}{3} \bar{g}_{\mu\nu} \) \xi + \O(\vp) \label{deltavp},
  }
where $\d_{\xi} h_{\mu\nu}$ receives corrections from the (higher
order) field redefinitions of the metric we have ignored. In
eq.~\eqref{deltavp} we recognize the gauge transformation that
characterizes a partially massless field~\eqref{pmsymmetry}. Thus, in
the perturbative metric formulation of the $SO(1,5)$ theory the local
$SO(2)$ rotations of the vielbeins are realized as the partially
massless symmetry of a massive graviton. This is reminiscent of how
diffeomorphisms in three-dimensional gravity correspond to gauge
transformations in the Chern-Simons formulation of the
theory~\cite{Witten:1988hc} (see also~\cite{Polyakov:1989dm}).  

Crucially, since the PM transformation is a consequence of the 
manifest $SO(2)$ invariance of the theory, we have indirectly
established that \emph{the PM symmetry exists to all orders in the 
fields}. The higher order terms we have neglected in
eqs.~\eqref{deltah2} and~\eqref{deltavp} guarantee the invariance of
the action order by order in $\vp_{\mu\nu}$. In particular, the cubic and higher order terms depend non-trivially on the gauge field $A$, thereby avoiding the obstructions to PM symmetry encountered in previous constructions~\cite{Zinoviev:2006im,deRham:2013wv,Joung:2014aba,Apolo:2016vkn}. Clearly, the non-linear vielbein formulation is the appropriate set up to verify that the partially
massless symmetry is realized to all orders. 

There is a price to pay for this non-linear realization of the
partially massless symmetry, however. While the quadratic 
action~\eqref{diagonalperturbativeaction} admits independent $PM$ and
$U(1)$ transformations, only the diagonal part of the $PM \times U(1)$
symmetry survives non-linearly. This can be readily established in the
vielbein formulation of the theory and is reflected in
eqs.~\eqref{deltaa3} and~\eqref{deltavp}. This implies that we lose
one of the first class constraints manifest in the quadratic theory
that is responsible for removing one of the helicity-0 modes from
either the massive graviton or the massless vector field. Thus,
without a proper Hamiltonian analysis, it is not clear whether all of
the helicity-0 modes decouple from the theory.\footnote{It may be
possible that the non-linear theory has an additional pair of second
class constraints which become first class upon linearization,
resulting in the $PM \times U(1)$ symmetry of the quadratic theory.}

Nevertheless, the fact that only the diagonal version of the
$PM \times U(1)$ gauge symmetry survives non-linearly does not mean
that the partially massless symmetry is trivially realized, e.g.~as in
the St{\" u}ckelberg trick. Rather, it is the precise form of the
transformation given in eq.~\eqref{deltavp}, along with all its
non-linear corrections, that guarantee invariance of the action under
the local $SO(2)$ transformation of the vielbeins. In particular, note that the diagonal part of $PM \times U(1)$ is sufficient to
maintain the relationship between the mass of the spin-2 field and the
cosmological constant, cf.~eq.~\eqref{higuchi}. Given that a similar
phenomenon is found in colored/charged theories of three-dimensional
gravity (at quadratic order)~\cite{Gwak:2015vfb,Gwak:2015jdo}, it is
natural to conjecture that the partially massless symmetry is a
generic feature of theories with charged metrics/vielbeins.

We conclude this section by pointing out another feature of the $SO(1,5)$ theory. Unlike the gauge
theory approach to Einstein and conformal gravity, where the gauge
groups used in the construction of the theory describe the global
symmetries of the latter, the $SO(1,5)$ theory does not admit a global
$SO(1,5)$ symmetry group. Indeed, while the quadratic
action~\eqref{diagonalperturbativeaction} \emph{does} admit such a
symmetry~\cite{Joung:2014aba}, at non-linear order only the $SO(1,4)$
symmetries of the de Sitter background survive. A similar result was
found in~\cite{Joung:2014aba} who studied the theory of a massless
spin-2 field and a partially massless graviton order by order in the
PM field. In that case, the loss of the $SO(1,5)$ symmetry at
non-linear order can be understood as a consequence of the obstruction
to the partially massless symmetry at higher
orders~\cite{Apolo:2016vkn}. In
contrast, the reason why the $SO(1,5)$ theory does not admit such a
global symmetry group is that only the diagonal part of the $PM \times
U(1)$ gauge symmetry survives non-linearly.


\section{Conclusions and outlook}
\label{se:conclusions}

In this paper we constructed a bimetric theory that realizes the
partially massless symmetry to all orders. The starting point was a
gauge theory based on the $SO(1,5)$ group, manifestly invariant under
its $SO(1,3)\times SO(2)$ subgroup, supplemented by additional
constraints. The resulting theory may be interpreted as an analytic
continuation of conformal gravity which is based on the $SO(2,4)$
group. The outcome is a bimetric theory with non-standard kinetic
terms and an additional vector field which, along with a specific
potential of bimetric gravity, render the theory invariant under local
$SO(2)$ transformations. Unlike conformal gravity, the linear spectrum
of the theory is free of ghost instabilities. More importantly, we
showed that in a perturbative formulation of the theory the local
$SO(2)$ symmetry is transmuted into the non-linear partially massless
symmetry of a massive spin-2 field.

One may expect a PM theory with an additional vector field to
propagate $2+4+2$ degrees of freedom corresponding to a massless
spin-2 graviton, a partially massless field, and the massless vector
field. While this is indeed the case at linear order, we have found
that only the diagonal $PM \times U(1)$ symmetry survives
non-linearly. This suggests that additional degrees of freedom may
propagate beyond linear order. Thus, it would be interesting to
analyze the constrained Hamiltonian of the theory and establish
whether the theory possesses enough constraints to propagate a total
of 8 degrees of freedom non-linearly. Otherwise, a helicity-0 mode
will be strongly coupled. 

We have also seen that our construction requires the complexification of the spin connection. As shown in Appendix~\ref{enhancedalgebra}, this spin connection is naturally associated with the
complexified Lorentz algebra, pointing towards an enlargement of the
starting $SO(1,5)$ group. One possibility is to consider a gauge
theory based on the complexified $\mathfrak{so}(1,5)$ algebra. We hope   
to report on this in the near future.

The structure of the kinetic terms used in the $SO(1,5)$ theory, where
the spin connection obeys the constraint eq.~\eqref{so15constraint},
also deserves further study. This is necessary to determine if
ghosts propagate non-linearly.  It would also be interesting to
determine whether a bimetric theory with such kinetic terms and a
generic bimetric potential, i.e.~one where the all of the $\b_n$
parameters in eq.~\eqref{metricaction} are arbitrary, leads to a
consistent theory. 

It is natural to expect that a generalization of the $SO(1,5)$ theory
leads to a non-abelian generalization of the partially massless
symmetry. If instead of $SO(1,5)$ one considers $SO(1,3+n)$, it is
possible to construct gauge theories for the $SO(1,3) \times SO(n)$
subgroups of the latter. These theories are characterized by $n$
vielbeins $e^a_{(i)}$ with $i = 1, \dots, n$, that transform
homogeneously under Lorentz transformations and as a vector under
$SO(n)$ rotations. In particular, if one imposes the symmetrization
condition~\eqref{vielbeinconstraint} for all the possible pairs of
vielbeins, the generalization of the action~\eqref{gaugeaction} leads
to
  \eq{
  I = &-\frac{M_p^2}{2} \int \e_{abcd} \bigg \{ \sum_{i=1}^n e^a_{(i)} \we e^b_{(i)}   \we R^{cd} - \frac{1}{2\ell^2}  \Big ( \sum_{i=1}^n e^a_{(i)} \we e^b_{(i)} \Big ) \we  \Big ( \sum_{i=1}^n e^c_{(i)} \we e^d_{(i)} \Big )   \bigg \}  \notag \\
  & - M_p^2 \ell^2\,\frac{\s^2}{2} \int \tr \( F \we \star\, F \),  \label{sonvielbeinaction}
  }
where $F = d A + A \we A$ is the field strength of the $SO(n)$ gauge field $A$. In analogy to the $SO(1,5)$ theory, the Hodge
dual in eq.~\eqref{sonvielbeinaction} is defined with respect to the
$SO(n)$ invariant metric given by $G_{\mu\nu} = \sum_{i=1}^n
e_{(i)}{}^a{}_\mu e_{(i) a\nu}$. The action~\eqref{sonvielbeinaction} is
manifestly $SO(1,3) \times SO(n)$ invariant provided that the spin
connection does not transform under $SO(n)$, as required from the
$\mathfrak{so}(1,3+n)$ algebra. 

Note, however, that the non-trivial part in the construction of the
$SO(1,5)$ theory is the constraint~\eqref{so15constraint} for the spin
connection. It is this constraint that forces us to make the latter
complex, contrary to what one would naively expect from gauging the
$\mathfrak{so}(1,5)$ algebra. Likewise, the non-trivial step that is necessary in
the construction of $SO(1,3+n)$ theories is the identification of an
appropriate constraint for the spin connection. The non-abelian nature
of $SO(n)$ makes this a difficult task that we leave for future study.


\section*{Acknowledgments}

It is a pleasure to thank Latham Boyle, Stanley Deser, Jelle Hartong,
Tomi Koivisto, Mikica Kocic, Anders Lundkvist, Angnis Schmidt-May,
Mikael von Strauss, Andrew Waldron, and Nico Wintergerst for comments
and fruitful discussions. This work was supported by a grant from the
Swedish Research Council. L.A. also thanks Bo Sundborg for
support. The perturbative expansion of the action in the metric
formulation of the theory was performed with the aid of the
Mathematica tensor package \emph{xAct}~\cite{xact,Nutma:2013zea}.


\appendix

\section{Enhanced algebra}
\label{enhancedalgebra}

In this section we consider a generalization of the $\mathfrak{so}(1,5)$ algebra that does not require the complexification of the spin connection introduced in Section~\ref{se:so15}. This algebra contains an additional antisymmetric generator $M_{ab}$, enlarging the algebra to a total of 21 generators, and satisfies
  \eq{
  [J_{ab}, J_{cd}] & = \eta_{ad} J_{bc} + \eta_{bc}J_{ad} - \eta_{ac} J_{bd} - \eta_{bd}J_{ac}, \label{JJ}\\
   [M_{ab}, J_{cd}] & = \eta_{ad} M_{bc} + \eta_{bc} M_{ad} - \eta_{ac} M_{bd} - \eta_{bd}M_{ac}, \label{MJ} \\
  [M_{ab}, M_{cd}] & = - \( \eta_{ad} J_{bc} + \eta_{bc}J_{ad} - \eta_{ac} J_{bd} - \eta_{bd}J_{ac} \), \label{MM} \\
  [J_{ab}, P^{(i)}_c] & = \d^{ij} \( \eta_{bc} P_{(j)a} - \eta_{ac} P_{(j)b} \), \\
  [M_{ab}, P^{(i)}_c] & = -\e^{ij} \( \eta_{bc} P_{(j)a} - \eta_{ac} P_{(j)b} \), \\
  [P^{(i)}_a, P^{(j)}_b] & = \e^{ij} \eta_{ab} D - \d^{ij} J_{ab}, \\
  [D, P^{(i)}_a] & = \e^{ij} P_{(j)a}, \label{DP}
  }
while all other commutators vanish. It is interesting to note that the eqs.~\eqref{JJ} --~\eqref{MM} can be obtained from a complexification of the Lorentz algebra, i.e.~by letting $J_{ab} \ra \tfrac{1}{2}( J_{ab} + iM_{ab})$. Thus, in some sense the complexification of the spin connection is unavoidable. More importantly, the algebra described by eqs.~\eqref{JJ} --~\eqref{DP} is not a Lie algebra since the Jacobi identity is not satisfied for all the possible combinations of the generators. Indeed, it is not difficult to check that $[M, [P^{(i)},P^{(j)}]] + \text{cyclic permutations} \ne 0$. Nevertheless, this algebra may be embedded into a larger Lie algebra where the Jacobi identity is satisfied for all the generators. One example of this is the complexification of the $\mathfrak{so}(1,5)$ algebra given by eqs.~\eqref{so15algebra1} --~\eqref{so15algebra4}.

Assuming that the algebra described by eqs.~\eqref{JJ} --~\eqref{DP} is the truncation of a consistent Lie algebra, the corresponding gauge field is now parametrized by
  \eq{
  \mathbb{A} = \frac{1}{2} \tau^{ab} J_{ab} + \frac{1}{2} \s^{ab} M_{ab} + \ell^{-1} e^a P^{(1)}_a + \ell^{-1} t^a P^{(2)}_a + A D + \dots
  }
If we focus on the $SO(1,3) \times SO(2)$ symmetries generated by this algebra we then find that $\tau^{ab}$ and $\s^{ab}$ transform as
  \eq{
  \d_{\l} \tau^{ab} &= \D_{\tau} \ll^{ab}, \qquad \d_{\l} \s^{ab} = \s^{a}{}_{c} \ll^{cb} + \s^{b}{}_{c} \ll^{a c}, \label{deltaomega3}
  }
where $\D_{\tau}$ is the covariant derivative with respect to the Lorentz connection $\tau^{ab}$. On the other hand the transformation of the vielbeins and the vector under $SO(1,3) \times SO(2)$ transformations remain unchanged. Note that the transformation of the connections $\tau^{ab}$ and $\s^{ab}$ given in eq.~\eqref{deltaomega3} is the same transformation inferred from the complexification of the spin connection, cf.~eq.~\eqref{deltaomega2}.

Let us now consider the field strengths associated to each of the generators of the enhanced algebra. We find,
  \eq{
        \mathbb{F}^{ab}_M  = & \D_{\tau} \s^{ab}, \\
  \mathbb{F}^{ab}_J = & \,\frac{1}{2} \( R_{\tau}^{ab} - \s^{a}{}_c \we \s^{cb} - \ell^{-2}\, e^a \we e^b - \ell^{-2}\, t^a \we t^b  \),  \label{enhancedFJ} \\
  \mathbb{F}^{a}_{P^{(1)}} = & \,\ell^{-1} \( \D_{\tau} e^a - \s^{a}{}_{b} \we t^b  +  A \we t^a \),  \\
  \mathbb{F}^{a}_{P^{(2)}} = & \,\ell^{-1} \( \D_{\tau} t^a  + \s^{a}{}_b \we e^b-  A \we e^a \),   \\
  \mathbb{F}_D = & \,d A + \ell^{-2} t^a \we e_a,
  }
where $R_{\tau}^{ab} = d \tau^{ab} + \tau^{a}{}_c \we \tau^{cb}$. In particular, the first two terms in eq.~\eqref{enhancedFJ} reproduce the real part of the complex curvature $R_{\om}^{ab} = d\om^{ab} + \om^{a}{}_c \we \om^{cb}$ that directly contributes to the action, see eq.~\eqref{realRw} . Also note that the constraint 
  \eq{
  \mathbb{F}^{a}_{P^{(1)}} + i  \mathbb{F}^{a}_{P^{(2)}}  = 0,
  }
reproduces the constraint given in eq.~\eqref{so15constraint} for the complex vielbein $\om^{ab} = \tau^{ab} + i \s^{ab}$. Furthermore, it is not difficult to check that all these curvatures transform homogeneously under local $SO(1,3) \times SO(2)$ transformations and that, in fact, the $\mathbb{F}^{ab}_M$, $\mathbb{F}^{ab}_J$, and $\mathbb{F}_D$ curvatures are left invariant under $SO(2)$ rotations. Thus the enhanced algebra presented in this section reproduces the desired curvatures, constraints, and transformations properties deduced from the complexification of the spin connection introduced in Section~\ref{se:so15}. Let us conclude by noting that the additional curvature term $\mathbb{F}^{ab}_M$ is important in the construction of the action. Indeed, this term guarantees that the higher derivative term $\int R_{\om}^{ab} \we R_{\om}^{cd} \e_{abcd}$, which is part of the \emph{gauge} theory action given in eq.~\eqref{gaugeaction}, is a topological invariant that does not contribute to the \emph{vielbein} action~\eqref{vielbeinaction}.



\ifprstyle
	\bibliographystyle{apsrev4-1}
\else
	\bibliographystyle{utphys2}
\fi

\bibliography{so15}



\end{document}